# MIMO Architectures for Efficient Communication in Space

Richard J. Barton, member
NASA Johnson Space Center
Houston, TX, USA
richard.j.barton@nasa.gov

*Abstract*—The characteristics of a space-to-space multiple-input, multiple-output (MIMO) communication channel that distinguish it from terrestrial MIMO communication channels are summarized and discussed primarily from an information-theoretic viewpoint. The implications of these characteristics for the design and application of future space-based communication systems are also discussed, and it is shown that in general, either energy-efficient or spectrally-efficient communication in space can only be achieved using a distributed MIMO architecture.

I. INTRODUCTION

In this paper, we consider distributed multiple-input, multiple-output (MIMO) communication in the context of a space-to-space communication system. Space-to-ground communication systems can be expected to have many of the same characteristics, but the distinguishing characteristics of space MIMO channels are most clearly revealed in the space-to-space environment. The three aspects of the space MIMO channel that distinguish it most dramatically from terrestrial applications are extremely long range, a free-space, line-of-sight (LOS) environment that is free of any significant scattering, and a need to design and build the communication system subject to severe constraints on the total antenna aperture area of the system.

A recent paper by the author [1] explored the characteristics and benefits of MIMO for space application heuristically, under the assumption that the multiplexing gain for a MIMO system in space scales with the number of antennas in exactly the same manner as that of a conventional terrestrial MIMO system operating in a rich scattering environment, with a simple constraint on total aperture area added to the equation. In a second paper which is still under review [2], the validity of this assumption is studied in a mathematically rigorous context and both upper and lower bounds on the ergodic capacity of a space MIMO channel are established in two related cases:

- In the first case, the channel is treated as random due to the fact that a fixed number of uniformly sized antennas are randomly distributed over a fixed spherical volume of space at both ends of the link. In this case, both the size of each individual antenna aperture and the transmitter power at each antenna scale linearly with the number of antennas in order to keep the total aperture area and total power fixed, and a single data stream is assumed transmitted from each antenna. This case is consistent with individual average transmitter power constraints at each antenna.

- In the second case, which is non-random, the channel is determined by a spherical region of space over which a fixed number of data streams are transmitted from an arbitrary number of antennas with total fixed aperture size that may be distributed aritrarily over the region. The power is scaled linearly over the data streams rather than the antenna elements, but no assumption is made regarding the number of individual antennas utilized to form the aperture or the size of each individual antenna. Hence, the total aperture area and the total power are still fixed, but the power distribution across the antenna elements and the size of each element are allowed to vary arbitrarily. Individual average transmitter power constraints at each antenna are inconsistent with this model.

Obviously, the second case is considerably more general than the first, but the second case provides a great deal of insight into what can be achieved in the first. In this paper, the results from both the earlier works are summarized and the implications for the design of energy-efficient deep space communication are discussed.

II. BACKGROUND

Throughout this paper, we are interested in comparing the performance of a conventional single-input, single-output (SISO) space communication system with an "equivalent" $M \times M$ MIMO system. For the SISO system, the channel is characterized by the *channel gain g* and the input *signal-to-noise ratio* (SNR) $\gamma$. These are given by

$$g = \frac{A_T A_R L}{\lambda^2 d^2}, \text{ and } \gamma = \frac{P}{BN_0},$$

respectively, where $B$ is the bandwidth of the transmitted signal, $P$ is the transmitted power, $A_T$ and $A_R$ are the effective areas of the (arbitrarily designated) transmitter and receiver antennas, respectively, $d$ is the range between the two antennas, $\lambda$ is the wavelength at the carrier frequency, $N_0$ is the power spectral density of the AWGN on the baseband equivalent (i.e., complex-valued) channel, and $L$ is a factor that represents the cumulative effect of additional unmodeled losses on the channel such as circuit losses, receiver noise figure, polarization losses, etc. Note that for this paper, since we are dealing with space communication links, we will always assume that $g \ll 1$.

For the equivalent MIMO channel, we also need to know the channel matrix **H**, which is given by

$$\mathbf{H} = \begin{bmatrix} h_{11} & h_{12} & \cdots & h_{1M} \\ h_{21} & h_{22} & \cdots & h_{2M} \\ \vdots & \vdots & \cdots & \vdots \\ h_{M1} & h_{M2} & \cdots & h_{MM} \end{bmatrix},$$

where $\{h_{ij}\}$, $i,j = 1,2,\ldots,M$, represent the complex-valued amplitude and phase couplings between receive antenna $i$ and transmit antenna $j$. For a terrestrial system, the equivalent MIMO channel is generally considered to be one in which the total transmit power, distributed across all of the individual transmitting antennas (often uniformly) is also given by $P$, but each of the transmitting and receiving antennas still has area $A_T$ or $A_R$, respectively. That is, the terrestrial MIMO system implicitly assumes the same total *equivalent isotropic radiated power* (EIRP) $P$ as the SISO system but with a receiver *array gain* of $M$.

For the space MIMO channel, if we really want to compare the performance of a MIMO architecture to a SISO architecture in any meaningful way, we must fix not only the total transmitted power but the total antenna aperture area at each end of the link. This is due to the fact that launch costs, which are directly related to mass and hence both system power and system antenna aperture area, are a very large (often dominant) factor in the total cost of deploying and operating a space communication system, and it means that we really need to scale the individual tranmsitter and receiver antennas to have area $A_T/M$ and $A_R/M$, respectively. This implies that there is no receiver array gain associated with a space MIMO system, and the total EIRP from the transmitter array actually *decreases* by a factor of $M$ with respect to the SISO system. At first blush, this would seem to imply that a space MIMO system can never be a good idea, and this may be the reason that MIMO antenna systems have not received more attention for space communication systems in the past. However, this is most definitely not the case. In fact, it turns out that for any desired level of spectral efficiency, energy-efficient communication in space can generally only be achieved with a MIMO communication architecture. Conversely, for any desired level of energy efficiency, spectrally-efficient communication in space can generally only be achieved with a MIMO communication architecture.

Consider first a conventional space communication system implemented by flying a single satellite with a single transceiver and a single antenna at each end of the link. For such a SISO link over an additive white Gaussian noise (AWGN) channel, the maximum achievable spectral efficiency in bits per second per Hertz (b/s/Hz) is given by [3,4]

$$\xi_1 = \log_2\left(1 + \frac{A_T A_R L P}{\lambda^2 d^2 B N_0}\right) = \log_2(1+\gamma g). \qquad (1)$$

Now consider the situation in which the same communication system is *fractionated* onto $M$ satellites at each end of the link, where each satellite is equipped with a single transceiver with power $P/M$, the satellites at one end all have single antennas with aperture area $A_T/M$, and the satellites at the other end all have single antennas with aperture area $A_R/M$. We assume that the satellites are distributed randomly over approximately spherical regions of space denoted by $\mathcal{V}_T$ and $\mathcal{V}_R$, respectively. This situation is illustrated in Figure 1 below.

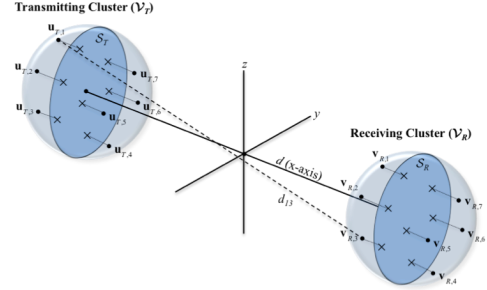

Fig. 1.   Space MIMO Example with *M*=7.

Assuming that the volumes of $\mathcal{V}_T$ and $\mathcal{V}_R$ are small compared to the distance $d$ and that the unmodeled losses are the same on each of the point-to-point channels between stations at each end of the link, the channel gain on each individual point-to-point channel will be well approximated by $g/M^2$ and the SNR on each channel will be given by $\gamma/M$. The remaining behavior of the channel is then determined by the structure of the channel matrix **H**, which will also be random and is discussed in more detail below. Whatever the actual structure of **H**, the maximum achievable spectral efficiency of the equivalent MIMO channel is then given by [5,6]

$$\begin{aligned}\xi_M(\mathbf{H}) &= \log_2\left(\det\left[\mathbf{I} + \frac{\gamma g}{M^3}\mathbf{H}\mathbf{H}^*\right]\right) \\ &= \sum_{i=1}^{M} \log_2\left(1 + \frac{\gamma g}{M^3}|v_i|^2\right),\end{aligned} \qquad (2)$$

where $\mathbf{H}^*$ is the complex-conjugate transpose of the matrix **H**, $\{|v_1|^2 \geq |v_2|^2 \geq \cdots \geq |v_M|^2\}$ are the eigenvalues of $\mathbf{H}\mathbf{H}^*$, and $\sum_{i=1}^{M}|v_i|^2 = M^2$.

Note that the spectral efficiency given by Equation (2) corresponds to channel capacity without channel side information at the transmitter [3]. That is, if the channel model is represented by

$$\mathbf{y} = \sqrt{\frac{g}{M^2}}\mathbf{H}\mathbf{x} + \mathbf{n}, \qquad (3)$$

where $\mathbf{n} \sim \mathcal{N}(\mathbf{0}, \mathbf{I})$ (i.e., complex Gaussian with mean zero and identity covariance matrix), then (2) represents the maximum achievable spectral efficiency under the constraint $E\{\mathbf{xx}^*\} = \frac{P}{M}\mathbf{I}$, which can be achieved using a codebook chosen from $\mathbf{x} \sim \mathcal{N}(\mathbf{0}, \frac{P}{M}\mathbf{I})$. We refer to this simply as the *uniform* spectral efficiency.

It turns out that for the space applications of interest, the uniform spectral efficiency does not actually depend on the true channel matrix $\mathbf{H}$ but on the version of the matrix determined by assuming that all antennas are projected onto the planar regions $\mathcal{S}_T$ and $\mathcal{S}_R$ depicted in Figure 1. Throughout this paper, we assume (without loss of much generality for the problem of interest) that $\mathcal{S} = \mathcal{S}_T = \mathcal{S}_R$ is a circle with radius $R$ and area $|\mathcal{S}| = \pi R^2$, and that $|\mathcal{S}|/\lambda d \geq 1$.

To derive all of the results presented in this paper, it is necessary to extend our channel model a bit. Towards that end, it is shown in [2] that the channel model given by Equation (3) is really just a constrained version of the more general channel model

$$y(\mathbf{v}) = \int_\mathcal{S} H(\mathbf{v}, \mathbf{u}) x(\mathbf{u}) d\mathbf{u} + n(\mathbf{v}), \quad (4)$$

where $\mathbf{u}, \mathbf{v} \in \mathcal{S}$, $n(\mathbf{v})$ is a complex Gaussian white noise process on $\mathcal{S}$ with power spectral density $N_0$,

$$H(\mathbf{v}, \mathbf{u}) = \sqrt{\frac{L}{\lambda^2 d^2}} e^{i\frac{2\pi}{\lambda d}\langle \mathbf{v}, \mathbf{u}\rangle},$$

and

$$x(\mathbf{u}) = \sum_{j=1}^{M} x_j f_j(\mathbf{u}),$$

represents symbols $\{x_j\}_{j=1}^{M}$ radiated from $M$ different arbitrarly defined *admissible distributed antennas* with transfer functions $\{f_j(\mathbf{u})\}_{j=1}^{M}$ that satisfy the following properties:

$$\int_\mathcal{S} f_i(\mathbf{u}) \overline{f_j}(\mathbf{u}) d\mathbf{u} = \delta_{ij}, \quad i, j = 1, 2, \ldots, M,$$

for each $i = 1, 2, \ldots, M$, there exists a set $\mathcal{A}_i \in \mathcal{S}$ such that

$$\mathcal{A}_i = \{\mathbf{u} \in \mathcal{S} : f_i(\mathbf{u}) \neq 0\}, \quad \mathcal{A} = \bigcup_{i=1}^{M} \mathcal{A}_i,$$

and either

$$|\mathcal{A}| = \int_\mathcal{S} I_\mathcal{A}(\mathbf{u}) d\mathbf{u} = A_T,$$

or

$$|\mathcal{A}| = \int_\mathcal{S} I_\mathcal{A}(\mathbf{u}) d\mathbf{u} = A_R,$$

depending on whether the distributed antennas are associated with the transmitting or receiving end of the link, respectively.

That is, in the more general case, the transmitted symbols may be radiated with varying intensity and phase from a continuum of points in $\mathcal{S}$ with aperture area $A_T$. The intensity and phase of the radiation are defined by the transfer functions of the distributed antennas, and the received signal $y(\mathbf{v})$ may be observed (through a set of distributed antennas) over a continuum of points in $\mathcal{S}$ with aperture area $A_R$.

If we let $\xi_M(\mathcal{S})$ represent the maximum achievable uniform spectral efficiency corresponding to Equation (4) for any choice of admissible distributed antennas, then it follows that any upper bound for $\xi_M(\mathcal{S})$ will also be an upper bound for $\xi_M(\mathbf{H})$ corresponding to Equation (3). Further properties of the relationship between $\xi_M(\mathbf{H})$ and $\xi_M(\mathcal{S})$ are explored in [2], and those properties are summarized and discussed below. However, the primary motivation for the results developed in [2] is to validate the use of the approximation

$$\xi_M(\mathbf{H}) = M \log_2\left(1 + \frac{\gamma g}{M^2}\right) \quad (5)$$

to explore the implications of MIMO communication for application in space that were derived in [1]. The results from [1] are also summarized and discussed below.

III. RESULTS AND DISCUSSION

All of the results in [1] were derived assuming that approximation (5) was valid. Those results are mainly concerned with exploring the relationship between the maximum achievable spectral efficiency for Problem (3), as represented by $\xi_M(\mathbf{H})$, and the minimum normalized energy per bit (or $E_b/N_0$) required to achieve that spectral efficiency, which is represented here by $\eta_M(\mathbf{H})$. Since $\eta_M(\mathbf{H})$ is the reciprocal of the number of bits that can be transmitted reliably over the channel per unit of normalized energy, this relationship captures the classical trade-off between spectral efficiency and energy efficiency on the channel. If Equation (5) is valid, then the relationship between $\xi_M(\mathbf{H})$ and $\eta_M(\mathbf{H})$ is given implicity by

$$\xi = \begin{cases} M \log_2\left(1 + \eta \xi \frac{g}{M^2}\right), & \eta \geq \eta_0, \\ 0, & \eta < \eta_0, \end{cases} \quad (6)$$

where $\eta_0$ is the *Shannon limit* for the channel, and we have dropped the dependence on *M* and **H** from the notation in order to simplify the expression. It is easy to show that Equation (6) implies that

$$\eta_0 = \frac{M}{g \log_2 e},$$

and that the corresponding *wide-band slope*, denoted here by $\omega_0$, (i.e., the slope of the solution to (6) as a function of $\tilde{\eta} = 10\log\eta$ evaluated at $\tilde{\eta}_0 = 10\log\eta_0$, see [7] and [8]) is given by

$$\omega_0 = M \frac{\log_2 e \cdot \ln 10}{5}.$$

The results from [1] are summarized in Figures 1-3 below. The value of the channel gain used to generate all of these figures is $g = 1$, which is just a normalization corresponding to using the received SNR rather than the input SNR in Equation (5). Figure 1 illustrates the solution to (6) over a wide range of values of η and *M*. Figure 2 illustrates the behavior of both the Shannon limit and the wide-band slope for the solution more clearly as a function of *M*, and Figure 3 illustrates the maximum achievable spectral efficiency and the corresponding number of required antennas as a function of η. If we define an energy-efficient architecture as one in which the minimum possible $E_b/N_0$ is utilized to achieve any particular spectral efficiency and a spectrally-efficient architecture as one in which the maximum possible spectral efficiency is achieved for any particular value of $E_b/N_0$, then it is clear from these three figures, that as a general rule, a MIMO architecture is required to achieve either energy-efficient or spectrally-efficient communication in space.

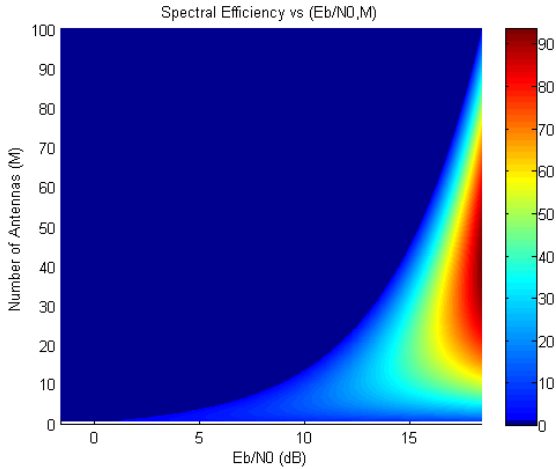

Fig. 1. Spectral efficiency for MIMO channel as a function of number of antennas and normalized received energy per bit.

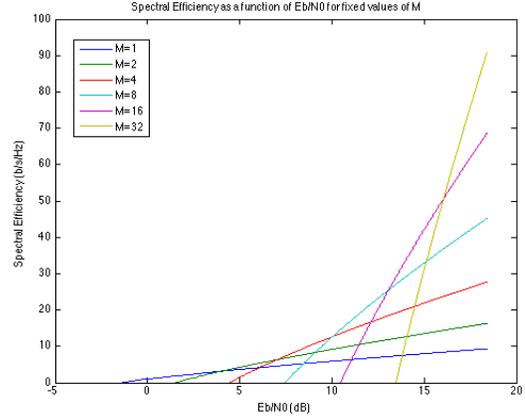

Fig. 2. Spectral efficiency vs. normalized received energy per bit for fixed antenna number on a MIMO link.

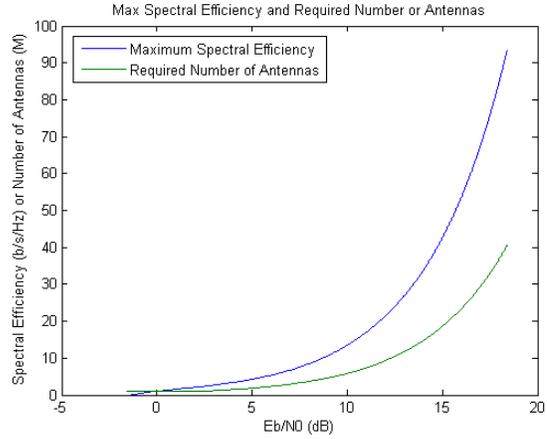

Fig. 3. Maximum value of spectral efficiency and corresponding number of required antennas on normalized MIMO link.

The following lemmas and their corollary are proven in [2] and taken together justify the use of the approximation given by Equation (5) as long as the area of the region $\mathcal{S}$ is sufficiently large. For both lemmas and the corollary, we let $\mathcal{M} = \left\lceil |\mathcal{S}|^2 / \lambda^2 d^2 \right\rceil$.

**Lemma 1.** For all matrices **H**, we have

$$\xi_M(\mathbf{H}) \leq \min\{M, \mathcal{M}\} \log_2\left(1 + \frac{\gamma g}{M \cdot \min\{M, \mathcal{M}\}}\right),$$

On the other hand, if the *M* transmit and receive nodes are independently and uniformly distributed over $\mathcal{S}$, then

$$E\{\xi_M(\mathbf{H})\} \geq \frac{\frac{M}{4}\log_2\left(1 + \frac{\lambda g}{2M^2}\right)}{\left(2 - \frac{1}{M}\right) + \frac{32}{9\pi}\left(M - 2 + \frac{1}{M}\right) / \frac{|\mathcal{S}|}{\lambda d}}.$$

**Lemma 2**. Let the kernel $H(\mathbf{v},\mathbf{u}): \mathcal{S} \to \mathcal{S}$ given in Equation (4) define a compact self-adjoint operator $\mathcal{H}$ mapping the Hilbert space $\mathcal{L}^2(\mathcal{S})$ of square-integrable functions on $\mathcal{S}$ into itself. As such it can be represented as

$$H(\mathbf{v},\mathbf{u}) = \sum_{n=1}^{\infty} \nu_n p_n(\mathbf{v}) \bar{p}_n(\mathbf{u}),$$

where $\{|\nu_1|,|\nu_2|,|\nu_3|,\ldots\}$ represents the sequence of eigenvalues for the operator $\mathcal{H}$ satisfying

$$|\nu_1| \geq |\nu_2| \geq |\nu_3| \geq \ldots,$$

$$\|\mathcal{H}\|^2 = \sum_{n=1}^{\infty} |\nu_n|^2 = \int_{\mathcal{S}} \int_{\mathcal{S}} |H(\mathbf{v},\mathbf{u})|^2 d\mathbf{u}\, d\mathbf{v} = \frac{L}{\lambda^2 d^2} |\mathcal{S}|^2,$$

and is a sequence of orthonormal functions that represent the eigenfunctions for $\mathcal{H}$. Furthermore, the set $\{p_n\}_{n=1}^{\infty}$ spans $\mathcal{L}^2(\mathcal{S})$ and either finitely many of the $\{\nu_n\}_{n=1}^{\infty}$ are nonzero or $|\nu_n| \to 0$ as $n \to \infty$. It follows that

$$\xi_M(\mathcal{S}) \geq \sum_{m=1}^{M} \log_2\left(1 + \frac{\gamma}{M} \cdot \frac{A_T A_R}{|\mathcal{S}|^2} |\nu_m|^2\right)$$

$$\approx \min\{M,\mathcal{M}\} \log_2\left(1 + \frac{\gamma g}{M\mathcal{M}}\right),$$

and this lower bound on $\xi_M(\mathcal{S})$ can be approached asymptotically using physically realizable collections of individual antenna elements.

**Corollary 1**. It follows from Lemmas 1 and 2 that

$$\xi_M(\mathcal{S}) \approx \min\{M,\mathcal{M}\} \log_2\left(1 + \frac{\gamma g}{M \cdot \min\{M,\mathcal{M}\}}\right).$$

Lemma 1 clearly implies that if the antennas are distributed randomly over a sufficiently large region $\mathcal{S}$, then

$$E\{\xi_M(\mathbf{H})\} \sim M \log_2\left(1 + \frac{\gamma g}{M^2}\right),$$

which is at least a partial justification for (5). Lemma 1 also shows that as the number of antennas in a space MIMO system grows, the antennas must be distributed over larger and larger regions in order to realize any gains from spatial multiplexing. Finally, Lemma 1 shows that for randomly placed antennas, there is a gap between the upper and lower bounds on $E\{\xi_M(\mathbf{H})\}$. On the one hand, the upper bound by itself implies that the maximum number of degrees of freedom on the channel is given by $\mathcal{M} \approx |\mathcal{S}|^2 / \lambda^2 d^2$. On the other hand, the lower bound by itself implies that performance may begin to degrade for any value of $M > |\mathcal{S}|/\lambda d$, which implies that the maximum number of degrees of freedom may be closer to $|\mathcal{S}|/\lambda d$.

The gap between the upper and lower bounds in Lemma 1 is closed by Corollary 1, which establishes that

$$\xi_M(\mathcal{S}) \approx \min\{M,\mathcal{M}\} \log_2\left(1 + \frac{\gamma g}{M \cdot \min\{M,\mathcal{M}\}}\right).$$

This essentially provides complete justification for Equation (5) in all cases, as long as we allow a slight generalization of Problem (3) that admits transmission of $M$ independent data streams over arbitrary linear combinations of a fixed set of distributed antennas subject only to a total power constraint. This also implies that the maximum number of degress of freedom on the channel is indeed given by $\mathcal{M} \approx |\mathcal{S}|^2 / \lambda^2 d^2$. Interestingly, an examination of the proof of Lemma 2 given in [2] shows that to achieve the maximum uniform spectral efficieny for the distributed-antenna space MIMO channel for a given number of independent data streams and a given radius of antenna array element distribution at both ends of the link, it is only necessary for the transmitter and receiver to exchange enough training information to establish and maintain a common coordinate system. As long as the transmitter and receiver both know where their own satellites are relative to the common coordinate system, then both can independently construct admissible antenna sets that will be asymptotically good.

REFERENCES

[1] R. J. Barton, "Distributed MIMO Communication Using Small Satellite Constellations," *Proceedings of the 2014 IEEE International Conference on Wireless for Space and Extreme Environments (WiSEE 2014)*, pp. 1-7, October 2014.

[2] R. J. Barton, "Properties of Space MIMO Communication Channels," arXiv:1601.03675 [cs.IT], 2016.

[3] T. M. Cover and J. A. Thomas, *Elements of Information Theory*, 2nd Edition, John Wiley & Sons, New York, 2006.

[4] J. G. Proakis, Digital Communications, 4th Edition, McGraw-Hill, New York, 2001.

[5] E. Telatar, "Capacity of Multi-Antenna Gaussian Channels," *European Transactions on Telecommunications*, vol. 10, no. 6, pp. 585–596, November 1999.

[6] E. Bigleiri, R. Calderbank, A. Constantinides, A. Goldsmith, A. Paulraj, H. V. Poor, *MIMO Wireless Communications*, Cambridge University Press, Cambridge, 2007.

[7] S. Verdu, "Spectral Efficiency in the Wideband Regime," *IEEE Transactions on Information Theory*, vol. 48, no. 6, pp. 1319-1343, June 2002.

[8] R. J. Barton, "Improved Estimation of the Spectral Efficiency Versus Energy-Per-Bit Tradeoff in the Wideband Regime," arXiv:1407.2518 [cs.IT], 2014.